\documentclass[graybox]{svmult}

\usepackage{amsmath}
\usepackage{amssymb}

\usepackage{type1cm}        
\usepackage{makeidx}         
\usepackage{graphicx}        
\usepackage{multicol}        
\usepackage[bottom]{footmisc}

\usepackage[numbers]{natbib}
\usepackage{newtxtext}       %
\usepackage{newtxmath}       

\usepackage{algorithm}

\usepackage{tikz}
\usetikzlibrary{backgrounds}
\usepackage{pgfplots}        
\usepgfplotslibrary{dateplot}

\pgfmathdeclarefunction{gauss}{2}{%
  \pgfmathparse{1/(#2*sqrt(2*pi))*exp(-((x-#1)^2)/(2*#2^2))}%
}

\makeindex             

\pgfplotsset{compat=1.16}
\begin{document}
\setcitestyle{square}

\title*{A fresh take on `Barker dynamics' for MCMC.}
\author{Max Hird, Samuel Livingstone \& Giacomo Zanella}
\institute{Max Hird \at Department of Statistical Science, University College London, Gower Street, London WC1E 6BT, UK. \email{max.hird.19@ucl.ac.uk}
\and Samuel Livingstone \at Department of Statistical Science, University College London, Gower Street, London WC1E 6BT, UK. \email{samuel.livingstone@ucl.ac.uk}
\and Giacomo Zanella \at Department of Decision Sciences, BIDSA and IGIER, Bocconi University, Via Roentgen 1, 20136 Milan, Italy. \email{giacomo.zanella@unibocconi.it}\vspace{3mm}\\
Preprint submitted to \emph{Proceedings of MCQMC2020} conference.
}
%
%
\maketitle

\abstract{We study a recently introduced gradient-based Markov chain Monte Carlo method based on `Barker dynamics'.  We provide a full derivation of the method from first principles, placing it within a wider class of continuous-time Markov jump processes.  We then evaluate the Barker approach numerically on a challenging ill-conditioned logistic regression example with imbalanced data, showing in particular that the algorithm is remarkably robust to irregularity (in this case a high degree of skew) in the target distribution.}

\section{Introduction}
\label{sec:intro}


For over half a century now Markov chain Monte Carlo has been used to sample from and compute expectations with respect to unnormalised probability distributions \cite{metropolis1953equation}. The idea is to construct a Markov chain for which a distribution of interest is invariant. Provided that the chain is $\pi$-irreducible and aperiodic (see e.g. \cite{roberts2004general}), then the distribution of $X_n$, the $n$th point in the chain, will approach the invariant distribution as $n\to\infty$, and ergodic averages from the chain can be used to approximate desired integrals.

Restricting attention to $\mathsf{X} \subset \mathbb{R}^d$, one way to confirm that a distribution on $\mathsf{X}$ with density $\pi(x)$ is invariant for a Markov chain with transition kernel $Q(x,A) := \int_A q(x,y)dy$ is to establish that the equation
\begin{equation} \label{eq:db}
\frac{\pi(y)q(y,x)}{\pi(x)q(x,y)} := t(x,y) =  1
\end{equation}
holds for all $x,y \in \mathsf{X}$ such that $\pi(x)q(x,y) > 0$, and that $\pi(y)q(y,x) = 0$ elsewhere. These are the well-known \textit{detailed balance} equations.  The celebrated Metropolis--Hastings algorithm \cite{metropolis1953equation,hastings1970monte} is built on the idea of coercing a Markov chain into having a specified invariant distribution. This is achieved through what will be called a \textit{balancing function} in this article.  Consider the scenario in which $\pi$ is not invariant for $Q$, meaning \eqref{eq:db} does not hold. A new kernel can be created which in fact \textit{does} satisfy equation \eqref{eq:db} by setting
\begin{equation} \label{eq:balanced_kernel}
p(x,y) := g(t(x,y))q(x,y),
\end{equation}
where $g(t)$ satisfies
\begin{equation} \label{eq:balancing_property}
    g(t) = tg(1/t)
\end{equation}
whenever $t>0$, and $g(0):=0$.  By noting that $\pi(x)q(x,y)t(x,y) = \pi(y)q(y,x)$ and $t(y,x) = 1/t(x,y)$, it is easily seen that 
\begin{align} \label{eq:db_p}
\pi(x)p(x,y) 
&= \pi(x)q(x,y)g(t(x,y)) \\ \nonumber
&= \pi(x)q(x,y)t(x,y)g(1/t(x,y))) \\ \nonumber
&= \pi(y)q(y,x)g(t(y,x)) \\ \nonumber
&= \pi(y)p(y,x)
\end{align}
as required.

The problem, however, with taking the above strategy is that there is no guarantee that $\int p(x,y) dy = 1$, in fact this is extremely unlikely to be the case. More steps must be taken, therefore, to create a Markov process. The Metropolis--Hastings solution is to restrict to balancing functions that satisfy $g(t) \leq 1$ for all $t \in[0, \infty)$. This ensures that the kernel $K(x,A) := \int_A p(x,y)dy$
satisfies $K(x,\mathsf{X}) \leq 1$. The remaining probability mass can then be found by simply adding a \textit{rejection step}, meaning that with probability $1-K(x,\mathsf{X})$ the chain remains at its current point $x$.

There is, however, another way to create a Markov process from $p(x,y)$, without resorting to the Metropolis--Hastings approach.  This consists of defining a continuous time Markov jump process in which jumps from the point $x$ occur with intensity
\begin{equation} \label{eq:jump_rate}
    \lambda(x) = \int p(x,y)dy,
\end{equation}
and the jump location $y$ is sampled from a distribution with density $p(x,y)/\lambda(x)$. The function $p(x,y)$ then describes the rate at which the process jumps from $x$ to $y$, and \eqref{eq:db_p} indicates that the process is $\pi$-invariant.  The challenge associated with this second approach is that the integral \eqref{eq:jump_rate} will often be intractable, meaning that simulating the process is not straightforward. Here we describe a solution that is outlined in the recent contribution \cite{livingstone2019barker},  through a judicious choice of $g$ and a suitable approximation to $t(x,y)$. It should of course be noted that in the case of finite $\mathsf{X}$ then \eqref{eq:jump_rate} becomes a sum, and so the process can be exactly simulated.  We do not consider this setting here, but direct the interested reader to \cite{power2019accelerated}, in which the approach is elegantly described, following earlier work in \cite{zanella2020informed}.

In the next section we discuss Barker's accept-reject rule, an early Markov chain Monte Carlo method from which we draw inspiration, before covering the general approach to the design of Markov jump processes with a prescribed invariant distribution in Section 3. It is here that we derive a Markov process that approximately preserves a given distribution, and show that the Barker balancing function is the only choice giving rise to such a process.  In Section \ref{sec:barkerprop} we reveal the Barker proposal scheme, in which this new process is used as a proposal mechanism within a Metropolis--Hastings algorithm. In Section \ref{sec:whyuse} we discuss the merits of using this new algorithm, by comparing it to suitable alternatives both theoretically and numerically, in the latter case using a challenging logistic regression example with imbalanced categorical data.

\section{Barker's rule and the Peskun ordering}
\label{sec:barker}

Readers who are familiar with the Metropolis--Hastings algorithm will naturally gravitate towards the choice of balancing function $g(t) = \min(1,t)$ in \eqref{eq:balanced_kernel}, resulting in the familiar Hastings acceptance probability
\begin{equation} \label{eq:mh_rule}
g_H(t(x,y)) = \min\left( 1, \frac{\pi(y)q(y,x)}{\pi(x)q(x,y)} \right).
\end{equation}
It should be noted, however, that several other choices of $g$ are possible.  One alternative proposed by Barker in \cite{barker1965monte} is $g(t) = t/(1+t)$, resulting in the acceptance probability
\begin{equation} \label{eq:barker_rule}
g_B(t(x,y)) = \frac{\pi(y)q(y,x)}{\pi(x)q(x,y) + \pi(y)q(y,x)} 
\end{equation}
after multiplying the numerator and denominator by $\pi(x)q(x,y)$.  In the case $q(x,y) = q(y,x)$ this further reduces to $\pi(y)/(\pi(x) + \pi(y))$. Note that both $g_H$ and $g_B$ satisfy $g \leq 1$.

The reason that $g_H$ is preferred to $g_B$ in the context of the Metropolis--Hastings algorithm is due to the work of Peskun \cite{peskun1973optimum} and Tierney \cite{tierney1998note}, which established that for the same choice of $q(x,y)$ the acceptance rate $g_H$ will result in Markov chains that produce ergodic averages with smallest asymptotic variance.  A key part of the argument is that $g_H$ will maximise the probability of moving from $x$ to $y$, for any $y \neq x$.  When comparing \eqref{eq:mh_rule} and \eqref{eq:barker_rule} this is easy to see, as
$$
g_B(t(x,y)) = \frac{\pi(y)q(y,x)}{\pi(x)q(x,y) + \pi(y)q(y,x)} \leq \frac{\pi(y)q(y,x)}{\pi(x)q(x,y)}
$$
whenever $\pi(x)q(x,y) > 0$. Combining with the fact that $g_B \leq 1$ gives that $g_B \leq g_H$ for every value of $t(x,y)$. 

It is important to emphasise, however, that the above discussion and conclusions about the optimality of $g_H$ are confined to the scenario in which \eqref{eq:balanced_kernel} is used to create a Metropolis--Hastings algorithm.  It no longer applies to the setting in which the function $p(x,y)$ is used to define a Markov jump process with transition rates given by \eqref{eq:jump_rate}. Furthermore, in this case the stipulation that $g \leq 1$ is not required.  This presents an opportunity to consider not just $g_H$ but also alternatives such as $g_B$ and others when designing jump processes of the type described in Section \ref{sec:intro}.


\section{Designing jump processes through a balancing function}
\label{sec:jumpprocess}


The dynamics of a jump process for which transitions from $x$ to $y$ occur at the rate $p(x,y)$ are as follows: if the current point at time $t$ is $X_t = x \in \mathsf{X}$, the process will remain at $x$ for an exponentially distributed period of time $\tau \sim \text{Exp}(\lambda(x))$, with $\lambda(x)$ defined in \eqref{eq:jump_rate}, before moving to the next point using the Markov `jump' kernel $J$, defined for any event $A$ as \begin{equation} \label{eq:jump_kernel}
    J(x,A) := \int_A \frac{p(x,y)}{\lambda(x)}dy.
\end{equation}  
In order to simulate such a process, we must therefore be able to compute $\lambda(x) := \int p(x,y)dy$, and also to simulate from $J(x,\cdot)$, for any $x \in \mathsf{X}$.

Note, however, that if $\lambda(x)$ were constant, we could simply use the jump kernel $J$ directly and simulate a discrete-time Markov chain. To see this, note that in general the jump kernel will have invariant density proportional to $\lambda(x)\pi(x)$, as the equation
$$
\lambda(x)\pi(x)\frac{p(x,y)}{\lambda(x)} = \lambda(y)\pi(y)\frac{p(y,x)}{\lambda(y)}
$$
simplifies to $\pi(x)p(x,y) = \pi(y)p(y,x)$, which holds by design. If $\lambda(x) = \lambda$ then the above equation simply shows that $J$ is $\pi$-reversible. We can therefore either simulate the continuous-time process with constant jump rate $\lambda$, or just ignore this step and take $J$ as the kernel of a discrete-time Markov chain.  In Subsection \ref{subsec:tractable}, we show that making a careful approximation to $t(x,y)$ allows just such a constant jump rate process to be found. 



\subsection{Tractability through a 1st order approximation of $t(x,y)$}
\label{subsec:tractable}


Recalling that $p(x,y) = g(t(x,y))q(x,y)$, the jump rate $\lambda(x)$ is the integral
$$
\int g(t(x,y))q(x,y)dy,
$$
which in general will not be tractable. A natural starting point for simplifying the problem is to restrict to the family of transition densities for which $q(x,y) = q(y,x)$, and further to the random walk case $q(x,y) = q(y-x)$. When this choice is made then $t(x,y) = \pi(y)/\pi(x)$. Restricting for now to $\mathsf{X} \subset \mathbb{R}$, a first order approximation of this ratio can be constructed using a Taylor series expansion as
$$
\pi(y)/\pi(x) = \exp\{\log\pi(y) - \log\pi(x)\} \approx \exp\{(y-x)\nabla\log\pi(x)\}
$$
for $y$ suitably close to $x$.  The purpose of using this approximation is that $y$ now only enters the expression through the difference term $z:= y-x$, and furthermore it holds that
\begin{equation} \label{eq:1storder_ratio}
t^*_x(z) := e^{z\nabla\log\pi(x)} = 1/e^{-z\nabla\log\pi(x)} = 1/t^*_x(-z).
\end{equation}
Since $q(x,y) = q(z)$ we can therefore express the entire integral as
$$
\lambda^*(x) := \int_{-\infty}^\infty g(t^*_x(z))q(z)dz.
$$
By first writing $\lambda^*(x)$ as the sum of two integrals over the disjoint regions $(\infty,0]$ and $[0,\infty)$, then switching the limits of integration through a change of variables in the first of these and finally re-combining, we arrive at the expression
$$
\begin{aligned}
\lambda^*(x) 
&= 
\int_{-\infty}^0 g(t^*_x(z))q(z)dz + \int_0^{\infty} g(t^*_x(z))q(z)dz \\
&= \int_0^{\infty} \left[ g(t^*_x(-z))q(-z) + g(t^*_x(z))q(z) \right]dz \\
&= \int_0^{\infty} \left[ g(t^*_x(-z)) + g(t^*_x(z)) \right]q(z)dz,
\end{aligned}
$$
where the last line follows from the fact that $q(z) = q(-z)$.
Using \eqref{eq:1storder_ratio} and then the balancing property \eqref{eq:balancing_property} reveals that
$$
\begin{aligned}
g(t^*_x(-z))
&= g(1/t^*_x(z))
= g(t^*_x(z))/t^*_x(z),
\end{aligned}
$$
meaning that setting $t^*_x(z):= t^*$ the term in square brackets inside the integral can be written $(1+1/t^*)g(t^*)$.  Note that if this expression were in fact equal to a constant, then $\lambda^*(x)$ would become tractable, and furthermore it would not depend on $x$. The Barker rule is the \textit{unique} (up to constant multiple) choice of balancing function for which this property holds. To see this, note that for any $c \neq 0$
$$
(1+1/t^*)g(t^*) = c \iff g(t^*) = \frac{c}{1+1/t^*}.
$$
Setting $c = 1$ and multiplying by $t^*/t^*$ reveals the choice $g_B(t^*) = t^*/(1+t^*)$, and furthermore 
$$
\lambda^*(x) = \int_0^\infty q(z)dz = \frac{1}{2},
$$
using the facts that $q(z) = q(-z)$ and $\int q(z)dz = 1$. In fact the choice of $c$ is irrelevant here as it simply acts as a constant multiple to the jump rate and does not enter into the jump kernel expression.  We refer to the resulting Markov process as \emph{Barker dynamics}.

\subsection{A skew-symmetric Markov transition kernel}
\label{subsec:skew}

The family of \textit{skew-symmetric} distributions on $\mathbb{R}$  has densities of the form
\begin{equation}\label{eq:skew_distr}
2F(\beta z) \phi(z),
\end{equation}
where $\phi$ is a symmetric probability density function, $F$ is a cumulative distribution function such that $F(0) = 1/2$ and $F'$ is a symmetric density, and $\beta \in \mathbb{R}$ \cite{azzalini2012some}. Choosing $\beta > 0$ induces positive skew and vice versa (setting $\beta = 0$ means no skew is induced). 
In fact $\beta z$ can be replaced with more general functions of $z$, but the above suffices for our needs.

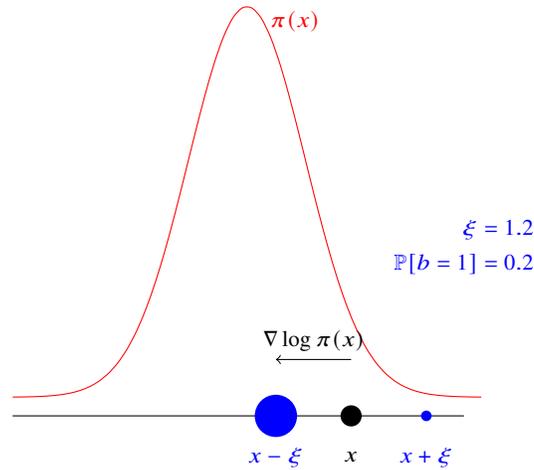
\begin{figure}[ht]
\begin{center}

\begin{tikzpicture}[scale=1]

\begin{axis}[every axis plot post/.append style={
  mark=none,domain=-4:4,samples=50,smooth},
  hide axis,
  enlargelimits=upper,
  ticks=none]
  
\addplot[color=red]{gauss(0,1)};
\end{axis}

\node[color=red] at (3.75,5) {$\pi(x)$};

\draw (0,-0.25) -- (6,-0.25);

\node[color=blue] at (6,2) 
{
$ 
\begin{aligned} 
\xi &= 1.2 \\ 
\mathbb{P}[b=1] &= 0.2 
\end{aligned} 
$
};

\fill (4.5,-0.25) circle (4pt);
\node[scale = 1] at (4.5,-0.8) {$x$};

\fill[color=blue] (3.5,-0.25) circle (8pt);
\node[color=blue, scale = 1] at (3.5,-0.8) {$x - \xi$};

\fill[color=blue] (5.5,-0.25) circle (2pt);
\node[color=blue, scale = 1] at (5.5,-0.8) {$x + \xi$};

\draw[->] (4.5,0.5) -- (3.5,0.5) node[pos=.5,sloped,above, scale = 1] {$\nabla\log\pi(x)$};

\end{tikzpicture}

\caption{A diagram of a typical draw from the transition kernel \eqref{eq:barker_transition} using the algorithm outlined in Subsection \ref{subsec:skew}. The black ball $x$ is the current state, and the sizes of the blue balls indicate the probability of moving to that point, given that the innovation drawn in step 1 is $\xi = 1.2$. A move in the direction of the gradient is clearly more probable.}
\label{fig:1dsketch}

\end{center}
\end{figure}

The jump kernel \eqref{eq:jump_kernel} with symmetric choice of $q$, the approximation $t^*$ in \eqref{eq:1storder_ratio} and Barker balancing function $g_B$ leads to the Markov kernel
\begin{equation} \label{eq:barker_transition}
J^*(x,A) := \int_A 2 g_B(\exp\{(y-x)\nabla\log\pi(x)\}) q(y-x) dy
\end{equation}
for any event $A$.  Writing $F_L(z) := 1/(1+e^{-z})$, the cumulative distribution function of the logistic distribution, and noting that $g_B(e^z) = F_L(z)$, the associated transition density can be written
$$
j^*(x,x+z) = 2F_L(\beta_x z) q(z)
$$
where 
$$
\beta_x := \nabla\log\pi(x).
$$
We see, therefore, that the resulting transition is \textit{skew-symmetric}, with the level of skew at the current state $x$ determined by $\nabla\log\pi(x)$.  Because of this, a convenient algorithm for drawing samples from this transition kernel exists, and consists of the following:
\begin{enumerate}
    \item Draw $\xi \sim q(\cdot)$
    \item Set $b = 1$ with probability $F_L(\beta_x \xi)$, otherwise set $b = -1$
    \item Set $z = b\xi$
    \item Return $x + z$.
\end{enumerate}
The resulting draw is from the kernel $J^*(x,\cdot)$. To see this, note that the probability density associated with any $z$ is
$$
j^*(x,x+z) = q(z)F_L(\beta_x z) + q(-z)(1-F_L(-\beta_x z)),
$$
which gives the density associated with either drawing $z$ and setting $b = 1$ or drawing $-z$ and setting $b = -1$.  After noting that $q(z) = q(-z)$ and $1 - F_L(-\beta_x z) = F_L(\beta_x z)$ by the symmetry of the logistic distribution, this simplifies to
$$
j^*(x,x+z) = 2F_L(\beta_x z)q(z)
$$
as required. Figure \ref{fig:1dsketch} illustrates the inner workings of such a transition.  It is natural to consider whether other choices of skewing function derived from other balancing functions can be used to produce such a Markov transition. It is shown in Appendix F of \cite{livingstone2019barker}, however, this is not possible, more precisely it is shown that $g_B$ is the unique choice of balancing function leading to a skew-symmetric transition kernel when the first order approximation $t^*(x,y)$ is used in place of $t(x,y)$. This is in fact evident from the calculations of Subsection \ref{subsec:tractable}.

\section{The Barker proposal in $\mathsf{X} \subset \mathbb{R}^d$}
\label{sec:barkerprop}

The culmination of Section \ref{sec:jumpprocess} is a Markov transition kernel \eqref{eq:barker_transition} and an algorithm in Subsection \ref{subsec:skew} to draw samples from this kernel. Note, however, that this transition kernel will not in general have equilibrium distribution $\pi$, owing to the 1st order approximation used in Subsection \eqref{subsec:tractable}.  In some cases it might be reasonable to simply ignore this fact and use the method regardless, in the hope that any approximation error is small (the authors will discuss this approach in forthcoming work). The resolution we will adopt here, however, is to use the transition as a proposal within a Metropolis--Hastings algorithm.  Note that the transition density can be written $j^*(x,y) \propto q(y-x)/(1+e^{(x-y)\nabla\log\pi(x)})$ with $q(y-x)=q(x-y)$, meaning that the Metropolis--Hastings acceptance probability becomes
$$
\alpha_1(x,y) = \min \left( 1, \frac{\pi(y)(1+e^{(x-y)\nabla\log\pi(x)})}{\pi(x)(1+e^{(y-x)\nabla\log\pi(y)})} \right).
$$

We have also restricted attention thus far to the one-dimensional setting, as extending to a $d$-dimensional transition kernel for $d>1$ can be done in many different ways.  It is natural to consider as a starting point a $d$-dimensional symmetric and centred density $q$. There are, however, many different ways to introduce the \textit{skewing} mechanism into a $d$-dimensional distribution, which is done in one dimension through the variable $b \in \{-1,1\}$. We consider two here, which we believe to be natural generalisations, and of which one is in fact clearly preferable to the other.  The first is to simply introduce the same variable $b$, and after drawing $\xi \sim q(\cdot)$, set $\mathbb{P}[b=1] = F_L(\beta_x^T\xi)$, where $\beta_x := \nabla\log\pi(x)$ as in Subsection \ref{subsec:skew}.  The only difference between this and the one-dimensional case is that now $\beta_x$ and $\xi$ are $d$-dimensional vectors, meaning the scalar product is replaced by an inner product. This procedure is a single global skewing of the initial symmetric distribution $q$.

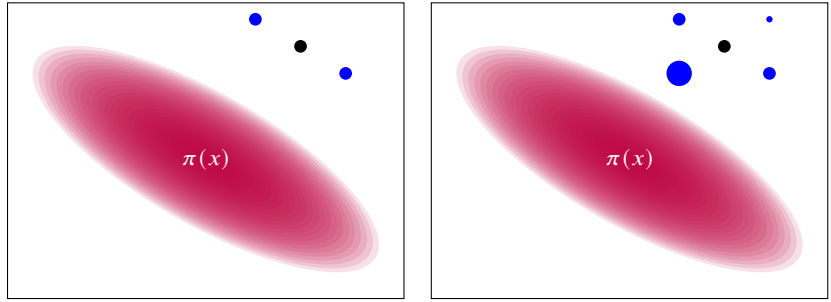
\begin{figure}[ht]
\begin{center}

\begin{tikzpicture}[framed, scale = 0.6]
\def\particles{(0,0)}
\foreach \point in \particles{
\foreach\i in {0,0.05,...,2.5} {
    \fill[opacity=\i*0.05,purple,rotate around={60:\point}] \point ellipse ({1-\i} and {3-3*\i});         
    }
} 
    
\fill (2.1,2.5) circle (4pt);
\fill [color=blue] (1.1, 3.1) circle (4pt);
\fill [color=blue] (3.1, 1.9) circle (4pt);
    
    
\node[color=white] at (0,0) {$\pi(x)$};

\end{tikzpicture}
\hspace{0.2cm}
\begin{tikzpicture}[framed, scale = 0.6]
\def\particles{(0,0)}
\foreach \point in \particles{
\foreach\i in {0,0.05,...,2.5} {
    \fill[opacity=\i*0.05,purple,rotate around={60:\point}] \point ellipse ({1-\i} and {3-3*\i});         
    }
} 
    
\fill (2.1,2.5) circle (4pt);
\fill [color=blue] (1.1, 3.1) circle (4pt);
\fill [color=blue] (3.1, 1.9) circle (4pt);
\fill [color=blue] (1.1, 1.9) circle (8pt);
\fill [color=blue] (3.1, 3.1) circle (2pt);
    
    
\node[color=white] at (0,0) {$\pi(x)$};
    
\end{tikzpicture}

\caption{A typical draw from the two different multi-dimensional transition kernels described in Section \ref{sec:barkerprop} when $d=2$. The black ball $x$ is the current state, and the sizes of the blue balls indicate the probability of moving to each candidate point after the initial innovation $\xi$ has been drawn. Using the first variant (left-hand side) only two moves are possible, neither of which move the chain closer to the high probability region of $\pi$. By contrast, using the second variant (right-hand side) $2^d$ moves are possible, and the most likely of these will move the chain in a favourable direction.}
\label{fig:2dsketch}
\end{center}
\end{figure}

It turns out, however, that a much more favourable approach is to skew each dimension individually. This involves defining $b \in \{-1,1\}^d$, and setting $\mathbb{P}[b_i = 1] = F_L(\beta_{x,i}\xi_i)$ for $i \in \{1,...,d\}$, where $\beta_{x,i} := \partial \log\pi(x)/\partial x_i$, the $i$th partial derivative of $\log\pi(x)$.  This approach allows a much more flexible level of skewing to be applied to the base distribution $q$. In fact, once the initial $\xi \sim q(\cdot)$ is drawn, the first approach only considers two possible candidate moves: $x+\xi$ and $x-\xi$. In a high dimensional setting it may not be that either of these candidate moves is particularly favourable in terms of helping the chain mix.  By contrast, the second approach allows for $2^d$ possible moves after $\xi$ has been sampled.  Figure \ref{fig:2dsketch} illustrates how this increased flexibility can result in much more favourable transitions.

One can make the comparison between the two approaches more concrete. In \cite{livingstone2019barker}, it is shown that the asymptotic variance of the first $d$-dimensional version of the Barker proposal will be at least half as large as that of a random walk Metropolis algorithm. As such, using scaling arguments based on limiting diffusion approximations, it can be shown that $O(d)$ iterations of the algorithm are needed to achieve estimates of a fixed level of precision as $d \to \infty$. By contrast, in the same work it is shown that only  $O(d^{1/3})$ iterations are needed for the second version to achieve the same goal. This is akin to the Metropolis-adjusted Langevin algorithm, another popular gradient-based Metropolis--Hastings algorithm (e.g. \cite{roberts1998optimal}). When referring to the Barker method in $d$-dimensions, from this point forward we will exclusively refer to the second approach described in this section.  A single transition of the resulting $d$-dimensional Metropolis--Hastings algorithm with current state $x$ is given below.

\begin{enumerate}
    \item Draw $\xi \sim q(\cdot)$
    \item For $i \in \{1,...,d\}$ set $b_i = 1$ with probability $(1+e^{-\beta_{x,i}\xi_i})^{-1}$, otherwise set $b_i=-1$, where $\beta_{x,i} := \partial \log\pi(x)/\partial x_i$
    \item Set $y := x + b \cdot \xi$, where $a \cdot b = (a_1b_1,a_2b_2,...,a_db_d)$ defines the element-wise product of two vectors $a = (a_1,...,a_d)$ and $b = (b_1,...,b_d)$ in $\mathbb{R}^d$
    \item Set the next state to be $y$ with probability
    $$
    \alpha_d(x,y) = \min\left( 1, \frac{\pi(y)}{\pi(x)} \prod_{i=1}^d \frac{1+e^{(x_i-y_i)\beta_{x,i}}}{1+e^{(y_i-x_i)\beta_{y,i}}} \right),
    $$
    otherwise remain at $x$.
\end{enumerate}
We note that the algorithm requires the same ingredients as MALA, and has the same computational cost per iteration, which is dominated by the calculation of the gradient and target distribution.  A simple function to run the Barker proposal in the R programming language is provided at \url{https://github.com/gzanella/barker}.

\section{Why use the Barker algorithm?}
\label{sec:whyuse}

Gradient-based MCMC methods are typically used because they perform well in high-dimensional settings. The Barker algorithm is no exception here, achieving the same $O(d^{-1/3})$ asymptotic efficiency as the popular Metropolis-adjusted Langevin algorithm (MALA) for suitably regular problems, where $d$ represents the dimension of the state space \cite{livingstone2019barker}.  The design of the Barker scheme, however, does differ from other gradient-based schemes such as MALA and Hamiltonian Monte Carlo (HMC).  In both of the latter well-known approaches the gradient is incorporated through a deterministic drift, which depends linearly on $\nabla\log\pi(x)$. In MALA, for example, if the current point is $x$ the proposal will be 
$$
y = x + \frac{h^2}{2}\nabla\log\pi(x) + h\xi,
$$
where $\xi \sim N(0,1)$ and $h > 0$. When the gradient is suitably regular and $h$ well-chosen this transition can be very desirable; for example if $\pi$ is Gaussian then the proposal becomes $y = (1-h^2/2)x + h\xi$, leading to dynamics in which the chain drifts towards the centre of the space very quickly provided that $h^2<2$. In the same setting, however, it is immediately clear that choosing $h^2 > 2$ will lead to undesirable behaviour.  The Barker proposal, by contrast, does not exhibit such a sharp cut-off between a good and bad choice of $h$ in this example.

The above case is indicative of a much more general phenomenon that is well-known to practitioners, namely that popular gradient-based methods often produce fast-mixing Markov chains on a particular class of problems and provided that the tuning parameters of the algorithm are well-chosen, but that this class of problems is smaller than ideal, and that performance degrades rapidly when a poor choice of tuning parameters is made.  This phenomenon is not only restricted to settings in which the MALA proposal becomes unstable (as in the Gaussian case), and means that it is also often difficult to tune the methods adaptively during the course of the simulation, an issue that is discussed in \cite{livingstone2019barker}.  In that work the authors focus on characterising \emph{robustness to tuning}, providing a mathematical argument to show that for MALA and HMC performance is much more sensitive to the choice of proposal tuning parameters than for the Barker proposal.



\subsection{Skewed target distributions}

One scenario in which gradient-based algorithms can perform poorly is when the distribution of interest $\pi$ exhibits considerable skew. To explore this phenomenon we first consider a simple one-dimensional model problem, before performing a more comprehensive numerical study on a challenging ill-conditioned logistic regression example.  We will show that in both of these cases the Barker algorithm is considerably more robust to the level of skewness exhibited than other gradient-based schemes.  In essence the challenge is that the gradient near the mode will diverge with the skewness of the distribution, causing pathologies in gradient-based proposals unless accounted for.

\emph{A model problem.}  Consider the family of skew-normal probability distributions $\pi_\eta$ on $\mathbb{R}$ indexed by a skewness parameter $\eta > 0$. A given member of the family will have density $\pi_\eta (z) := 2 \phi(z) \Phi(\eta z)$ where where $\phi$ and $\Phi$ are the density and cumulative distribution function of a standard normal distribution. Note that as $\eta$ increases so does the skewness and that $\pi_\eta$ becomes a truncated Gaussian truncated to be positive as $\eta\to\infty$. Take $x>0$ larger than the mode of $\pi_\eta$, and set $y=0$, noting that this implies $\text{sign}(\nabla \log \pi_\eta(x)) = -\text{sign}(\nabla \log \pi_\eta(y))$. The choice $y=0$ is important only for the limiting result, in reality algorithmic difficulties will occur for any point in the neighbourhood of zero for which the gradient is large and positive when $\eta \gg 0$.  For these choices, as $\eta \to \infty$ it holds that $\pi_\eta(x) \to 2\phi(x)$, $\pi_\eta(y) \to 1/\sqrt{2\pi}$, and $\nabla \log \pi_\eta(x) \to -x$, whereas $\nabla \log \pi_\eta(y)) \to \infty$ as the density becomes increasingly skewed.  Recall that the MALA proposal density is
$$
\log q_\eta^M (z_1, z_2) := -\frac{1}{2 h ^ 2}\left(z_2 - z_1 - \frac{h ^ 2}{2}\nabla \log \pi_\eta (z_1)\right) ^ 2 - \frac{1}{2}\log\left(2 \pi h ^ 2 \right).
$$
This implies that $\log q_\eta^M (y, x) \to -\infty$ as $\eta \to \infty$, whereas $\log q_\eta^M (x, y)$ remains finite. As a consequence the reverse move from $y$ to $x$ becomes increasingly unlikely as $\eta$ grows, causing the acceptance rate
$$
\alpha_\eta^M(x, y) := \text{min}\left(1, \frac{\pi_\eta (y) q_\eta ^ M (y ,x)}{\pi_\eta (x) q_\eta ^ M (x ,y)}\right)
$$
to become arbitrarily small, such that $\alpha_\eta^M(x,y) \to 0$ as $\eta \to \infty$. The Barker proposal density is
$$
\log q_\eta^B(z_1, z_2) := -\log\left(1 + \exp((z_1 - z_2)\nabla \log \pi_\eta(z_1)\right) + C
$$
for some finite constant $C$. Since $(y - x)$ and $\nabla \log \pi_\eta(y)$ have opposite signs, their product tends to $-\infty$ in the same limit, meaning $\log q_\eta^B(y, x) \to C$. 
The acceptance rate $\alpha_\eta^B(x,y)$ for the Barker algorithm therefore remains stable and converges to a positive value in the same limit. 

\begin{figure}[ht]
    \centering
    \includegraphics[width=11cm]{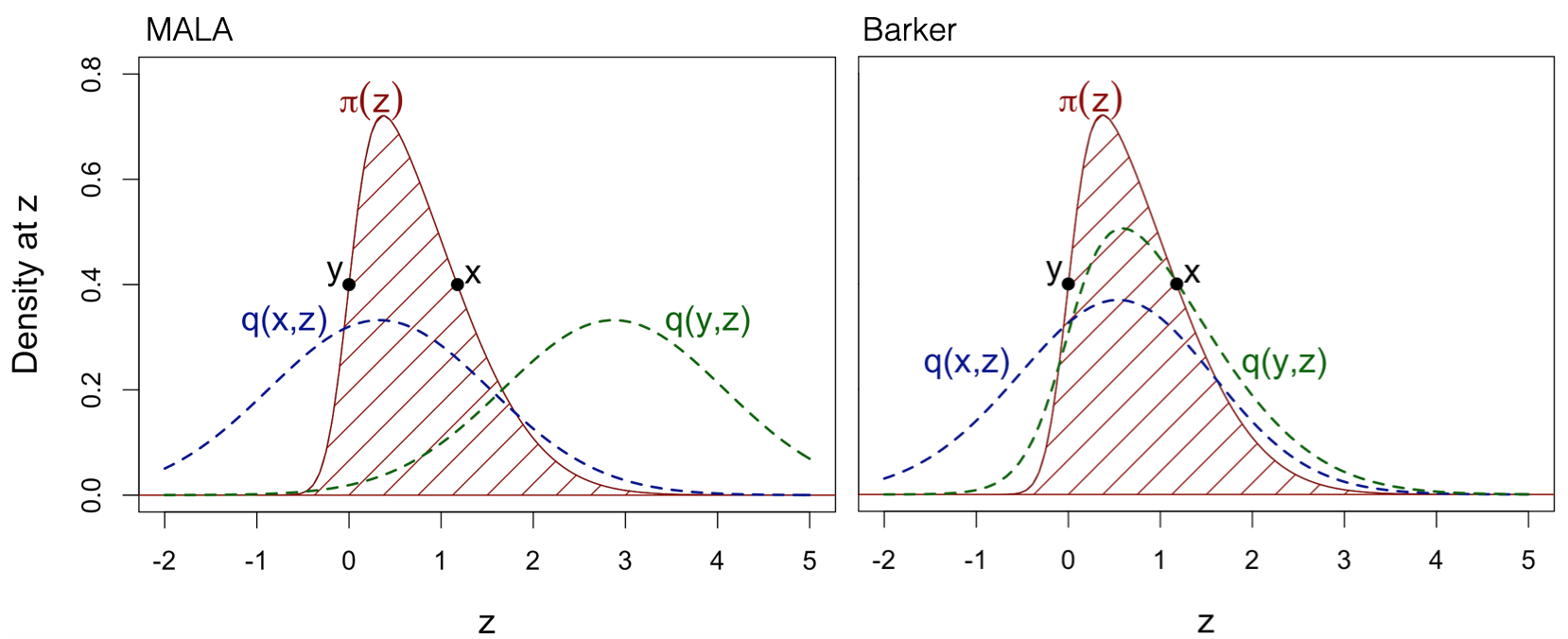}
    \caption{The forward (blue) and reverse (green) proposal densities associated with two points separated by a mode for an example target distribution that contains skew. In the MALA case the current point $x$ is quite unlikely under the reverse proposal density (green curve), whereas for the Barker algorithm this is not the case.}
    \label{fig:skew_example}
\end{figure}

Figure \ref{fig:skew_example} provides some more intuition for the contrasting behaviour between the two methods in this example.

\subsection{A logistic regression example with imbalanced data}

Skewed posterior distributions appear in many common modelling settings, but it is perhaps surprising that even seemingly simple logistic regression models can exhibit such a degree of skew that they pose a significant challenge to MCMC methods. This is despite the fact that the posterior distribution is strongly log-concave and the gradient is Lipschitz, meaning that several favourable results on the mixing properties of classical gradient-based algorithms can been established (e.g. \cite{durmus2017nonasymptotic,dalalyan2019user,dwivedi2018log}). 

We consider an example  using the arrythmia dataset from the UCI machine learning repository, available at \url{https://archive.ics.uci.edu/ml/datasets/arrhythmia}. The dataset consists of 452 observations of 279 different covariates. The modelling task is to detect the presence or absence of cardiac arrythmia.  The data presents a challenge as there are many imbalanced categorical covariates with only a few observations in certain categories. 

\begin{figure}[t]
    \centering
    \includegraphics[width=11cm]{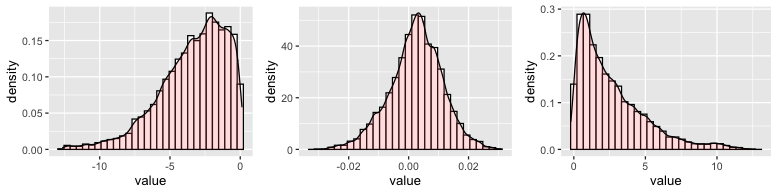}
    \caption{Example marginal distributions for selected covariates using the output of the Barker algorithm, illustrating varying degrees of skew in different dimensions. The plots are shown for the raw data.}
    \label{fig:histograms}
\end{figure}

The number of predictors compared to the size of the dataset makes the problem highly ill-conditioned. To combat this we selected 25 imbalanced covariates and 25 others, meaning 50 covariates in total for our problem.  The 25 imbalanced predictors were chosen from among the categorical covariates for which one category appeared two or fewer times in the dataset, whereas the remaining 25 were chosen from the remaining set.  Despite this pre-processing the problem is still highly ill-conditioned and the maximum likelihood estimator is undefined, making a Bayesian approach very natural for the problem. We also note that despite the reduced number of covariates the final problem is still of large enough dimension that simpler fitting methods will be ineffective, in line with the recommendations of \cite{chopin2017leave}.  The choice of dataset was inspired by \cite{johndrow2019mcmc}, in which the authors highlight that imbalanced categorical data can cause problems for Markov chain Monte Carlo methods. In this case the result is a logistic regression posterior distribution with a pronounced level of skewness in certain dimensions, as shown in Figure \ref{fig:histograms}.

For the Barker scheme we choose a Gaussian $q(\cdot)$ (although we note anecdotally that ongoing work suggests that other choices may be preferable).  With the goal of minimising the degree of hand-tuning needed for each algorithm, we used an adaptive approach to choosing algorithmic tuning parameters, precisely  Algorithm 4 of \cite{andrieu2008tutorial}, which consists of a Robbins--Monro scheme for learning a single global scale $\lambda$ and a covariance matrix $\Sigma$, which combine to form the pre-conditioning matrix $\lambda^2 \Sigma$. We set the Robbins--Monro learning rate at time $t$ to be $t^{-0.6}$.  The matrix $\Sigma$ can be dense or restricted to diagonal; the former allows correlations to be better navigated by the sampler, but the diagonal approach means less parameters must be learned during the simulation.  A weakly-informative independent Gaussian prior with zero mean and variance $25$ was chosen for each model parameter. It is also sometimes recommended in logistic regression problems to first \emph{standardise} the covariates, transforming each to have zero mean and unit variance. This can have the effect of making the posterior more regular and as a consequence the inference less challenging, but is not always done by practitioners. In our case the scales by which the covariates were standardised range from \textasciitilde0.05 to \textasciitilde32.

The above considerations led us to four different testing scenarios for each algorithm: dense $\Sigma$ with raw data, dense $\Sigma$ with standardised data, diagonal $\Sigma$ with raw data and diagonal $\Sigma$ with standardised data. For each of these scenarios we compared the Barker proposal scheme with MALA, a classical gradient-based alternative, as a simple illustration of the different patterns of behaviour that the two algorithms can exhibit.

\begin{figure}[ht]
    \centering
    \includegraphics[width=11cm]{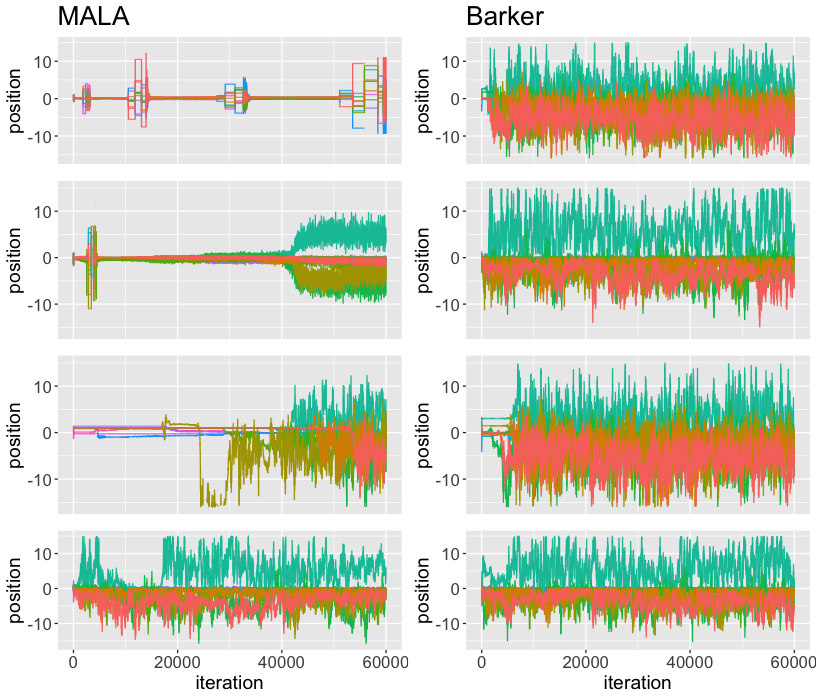}
    \caption{A selection of trace plots from the MALA and Barker algorithms for the logistic regression example. 1st row: raw data with dense $\Sigma$; 2nd row: standardised data with dense $\Sigma$; 3rd row: raw data with diagonal $\Sigma$; 4th row: standardised data with diagonal $\Sigma$.}
    \label{fig:traceplots}
\end{figure}

Trace plots showing the performance of the MALA and Barker algorithms in each scenario are shown in Figure \ref{fig:traceplots}.
It is immediately clear that MALA struggles to reach equilibrium in 3 out of 4 scenarios, only really performing reasonably when $\Sigma$ is diagonal and the data is standardised.  As expected, standardising the data aids performance, but it is perhaps surprising that the sampler also struggles in the dense $\Sigma$ setting. By comparison, visually the Barker algorithm behaves reasonably in all scenarios.  To evaluate the samplers at equilibrium and once the adaptation has stabilised, we examine effective sample sizes for each scenario in which equilibrium is visually reached after 30,000 iterations in Table \ref{tab:effective_sizes}.
\begin{table}[ht]
\caption{Minimum and median effective sample sizes (ESS) for the logistic regression example.}
\label{tab:effective_sizes}       
%
%
\begin{center}
\begin{tabular}{p{2.5cm}p{3.5cm}p{4cm}}
\hline\noalign{\smallskip}
Dataset & Algorithm & ESS (min., med.) \\
\noalign{\smallskip}\svhline\noalign{\smallskip}
Raw & Barker (dense)  & 38.82, 156.67 \\
Raw & Barker (diag.)  & 65.55, 164.67 \\
Raw & MALA (dense)   & n/a \\
Raw & MALA (diag.)   & n/a \\

Standardised & Barker (dense)  & 53.36, 98.44 \\
Standardised & Barker (diag.)  & 44.19, 101.51 \\
Standardised & MALA (dense)    & n/a \\
Standardised & MALA (diag.)    & 37.21, 87.14 \\
\noalign{\smallskip}\hline\noalign{\smallskip}
\end{tabular}
\end{center}
\end{table}
The effective sample sizes allow us to see that performance once equilibrium is reached is largely comparable between the two schemes once the scenario is favourable. The key strengths of the Barker approach in this example are its robustness to lack of standardisation and robustness to different adaptation strategies.

\section{Discussion}

We have given a pedagogical treatment of the Barker proposal scheme, a new gradient-based MCMC algorithm that we argue has some desirable features when compared to classical gradient-based alternatives, namely its robustness (in a very general sense).  There are numerous ways in which classical schemes such as MALA and HMC can be made more robust in different settings (e.g. \cite{roberts1996exponential,brosse2019tamed,livingstone2019kinetic,lu2017relativistic}), but these often introduce additional tuning parameters and can suffer from other issues, meaning that the quality of performance becomes very problem-specific. Another alternative approach are second-order methods that incorporate the Hessian of $\log\pi(x)$ in some way (e.g. \cite{girolami2011riemann,livingstone2014information}), but generally the cost of their implementation is large, and can grow cubically with dimension.  Based on the simplicity, scaling properties and robustness of the Barker proposal we argue that there are likely to be many realistic scenarios in which it proves useful, and in addition there is much room for the development of further algorithms within the general framework discussed in Section \ref{sec:jumpprocess}.


\bibliographystyle{spmpsci}
\bibliography{references}

\end{document}